\documentclass[aps,pra,twocolumn,superscriptaddress,floatfix,english]{revtex4-1}
\usepackage{graphicx}
\usepackage{multirow,bigdelim}
\usepackage{bm}
\usepackage{amsmath}
\usepackage{lipsum}
\usepackage[breaklinks=true,pdflang=en]{hyperref}
\usepackage{float}

\usepackage[english]{babel}
\usepackage[utf8]{inputenc}
\usepackage{blindtext}

\usepackage{xcolor}

\hypersetup{colorlinks=true,
pdftitle={1.5T Voigt spectroscopy paper},
pdfauthor={Francisco S. Ponciano-Ojeda},
linkcolor=blue,
citecolor=blue,
urlcolor=blue}

\begin{document}

\title{Absorption spectroscopy and Stokes polarimetry in a $^{87}$Rb vapour in the Voigt geometry with a 1.5~T external magnetic field}
\date{\today}

\author{Francisco S. Ponciano-Ojeda}
\email{francisco.s.ponciano-ojeda@durham.ac.uk}
\affiliation{Joint Quantum Centre (JQC) Durham-Newcastle, Department of Physics, Durham University, South Road, Durham, DH1 3LE, United Kingdom}
\author{Fraser D. Logue}
\affiliation{Joint Quantum Centre (JQC) Durham-Newcastle, Department of Physics, Durham University, South Road, Durham, DH1 3LE, United Kingdom}
\author{Ifan G. Hughes}
\affiliation{Joint Quantum Centre (JQC) Durham-Newcastle, Department of Physics, Durham University, South Road, Durham, DH1 3LE, United Kingdom}

\begin{abstract}
\textcolor{black}{This paper provides details of a} spectroscopic investigation of a thermal $^{87}$Rb atomic vapour.
\textcolor{black}{The experiment was conducted with an external magnetic field of 1.5~T in the Voigt geometry.}
\textcolor{black}{Very good quantitative agreement between experimental data and theory is found for all four Stokes parameters--with RMS errors of $\sim 1.5$\% in all cases.}
\textcolor{black}{From the fits to our experimental data a value for the magnetic field strength is extracted, along with the angle between the magnetic field and the polarisation of the light.}
\textcolor{black}{The effects of the cell window birefringence on the optical rotation signals are characterised.}
This allows us to carry out precise measurements at a high field strength and arbitrary geometries, allowing further development of possible areas of application for atomic magnetometers.
\end{abstract}

\maketitle


\section{Introduction}
\label{sec:intro}

Over the past decades atomic spectroscopy has cemented itself as one of the cornerstones of precision measurements.
From helping define the standards for time-keeping~\cite{Terrien1968,Campbell2011,Ludlow2015} to measurement of electromagnetic fields~\cite{Osterwalder1999, Thiele2015, Sedlacek2013,Kumar2017,Horsley2015,Horsley2016}, spectroscopic techniques have become key tools in high precision systems.
In recent years, together with the push for new technologies in other areas, this has seen the field of atom-based sensors burgeon.
Of particular interest are magnetic field sensors~\cite{Budker2007} which rely on the interaction of the nuclear and electronic spins with the external field and well-known spectroscopic signals.
Applications of such atomic sensors span a variety of disciplines, \textcolor{black}{ranging from explosives detection~\cite{Lee2006}; microfluidics~\cite{Xu2008}; medical imaging of soft tissues~\cite{Bison2009,Alem2015,Sander2012,Boto2017}; gyroscopes~\cite{Donley2009}; and measurements on spin-active solid-state systems~\cite{Arnold2016}.}
In order to improve and sustain the development of these technologies, atomic spectroscopy in the presence of external magnetic fields has become an area of wide interest~\cite{Sargsyan2015b,Arimondo2016,Whiting2017b,Si2020,Nyakango2020}. 
This in turn has given rise to a variety of experiments and optical devices~\cite{Weller2012b,Abel2009a,Kiefer2014,Zentile2015g,Zentile2015a,Rotondaro2015,Keaveney2016c,Keaveney2018b} that help demonstrate the depth of understanding of the physics involved in said interactions, furthering the reach and possible applications of this area.
Nevertheless, this in-depth understanding of the interactions between atoms and an externally applied magnetic field has primarily occurred for fields up to $\sim 1$~T.

The use of spectroscopic techniques in atomic systems has received less attention at the higher end ($>1$~T) of the range of field strengths for different reasons.
There is the matter of most methods used to obtain large fields being destructive in their nature, thus complicating the experimental reproducibility.
In such experiments, the Zeeman splitting of energy levels in alkali-metal atoms is used to observe field strengths of the order of tens/hundreds of Tesla via spectroscopic techniques~\cite{Gomez2014,Garn1966,Banasek2016}.
Non-destructive techniques for producing these large fields also exist, and this has enabled work in large pulsed magnetic fields up to $58$~T with similar alkali-atom systems~\cite{Ciampini2017a,George2017}.
Aside from the large Zeeman splitting produced at such large magnetic fields, there are other changes in the atoms that in turn allow additional effects to be observed.
Of these changes the most relevant is the decoupling of the nuclear and electronic spins in the atom, with the external magnetic field now being a common axis for the precession of both.
This is known as the hyperfine Paschen-Back regime~\cite{Umfer1992,Windholz1985,Windholz1988,Olsen2011,Weller2012a,Zentile2014a,Sargsyan2017} and, in contrast to the experiments at lower fields ($<10$~mT) -- where it is easily and directly measured~\cite{Budker2007} -- the Larmor precession frequency is not of interest as it can be too high for conventional electronic systems.
Rather, the characteristic absorption spectrum of the atomic system is a more straightforward way of obtaining information about the magnetic field.

Work has previously been done in atomic vapours of both low and high atomic densities~\cite{Siddons2008b,Weller2011a,Lange2020}, and other relevant magneto-optic effects have been the subject of extensive studies~\cite{Siddons2009a,Siddons2009b,Siddons2010,Weller2012}.
This atom-light interaction whilst in the presence of an external magnetic field has also been used in other systems such as~\cite{Whiting2015,Whiting2016a,Whiting2017,Whiting2017b}.
\textcolor{black}{For the majority of said work} the $\vec{k}$ vector of the laser beam used is parallel to the direction of the external magnetic field $\vec{B}$\textcolor{black}{--the Faraday geometry}. This gives rise to characteristic absorption spectra that are highly symmetric~\cite{Weller2012a,Sargsyan2014} and reflect only the magnitude of the magnetic field~\cite{Zentile2014a}.
However, there has been some work~\cite{Sargsyan2015d,Keaveney2019} that makes use of the less studied Voigt configuration where, rather than being parallel, the $\vec{k}$ vector of the light and the direction of the magnetic field $\vec{B}$ are perpendicular.
The change in geometry results in a change in the atomic transitions permitted by the selection rules \cite{Sargsyan2014b,Hakhumyan2020} and thus allows for information not only on the magnitude but also on the direction of the magnetic field to be experimentally observed.

\textcolor{black}{In this work we provide a detailed spectroscopic analysis of measurements taken in the Voigt geometry with an atomic vapour at a magnetic field strength of $1.5$~T.}
These measurements include optical rotation signals, expressed in terms of the Stokes parameters.
\textcolor{black}{Our theoretical model \textit{ElecSus} describes the atomic susceptibility~\cite{Zentile2015b,Keaveney2017a} and provides a means of fitting the experimental data, from which values of the magnetic field strength are extracted and compared against expected values.}
\textcolor{black}{From the fit we also extract the relative orientation of the magnetic field to a given known input light polarisation.}
Finally, we demonstrate a method to measure the birefringence of the cell windows by looking at small changes in the expected optical rotation signals produced by the atomic medium in the presence of a strong external magnetic field.

\textcolor{black}{The text continues in the following manner: section~\ref{sec:theory} gives a brief summary of the theory behind the model used in analysing our experimental data; the experimental methods to obtain these data are presented in section~\ref{sec:expt}; a discussion and comparison of our model and the experimental spectra appear in section~\ref{sec:spectra}.
The paper finishes with further elaboration upon the analysis of our experimental data in section~\ref{sec:birref} where} the optical rotation signals \textcolor{black}{are used} to extract information regarding the birefringence introduced by the cell windows.

\section{Theoretical model}
\label{sec:theory}

\textcolor{black}{In order to fit our experimental data we use \textit{ElecSus}.
The theory behind this software has been treated} in detail in refs.~\cite{Zentile2015b,Keaveney2017a}.
\textcolor{black}{For this work} we summarise \textcolor{black}{this theory and highlight the} points \textcolor{black}{of most relevance}.
We begin by assuming an atom-light system operating in the weak-probe regime~\cite{Smith2004,Sherlock2009}, although recent work~\cite{Lange2020} provides methods to generalise this into a strong-probe regime.
\textcolor{black}{The base for the model lies in using} the complex electric susceptibility \textcolor{black}{of the atomic medium}, $\chi(\Delta)$, as a function of the optical frequency detuning $\hbar\Delta = (\hbar\omega_{\rm{laser}} - \hbar\omega_{0})$ \textcolor{black}{near the alkali D-line resonances.}
\textcolor{black}{Here we define $\omega_{\rm{laser}}$ as the angular frequency of the laser and $\omega_{0}$ as} the angular frequency of the atomic transition.
\textcolor{black}{We also note that experimentally it is more common to use linear detuning, $\Delta/2\pi$}.
Using a matrix representation we construct the atomic states \textcolor{black}{for our system} in the uncoupled $m_L, m_S, m_I$ basis.
\textcolor{black}{We can then account for the }internal energy levels that come about as a result of the fine and hyperfine structure \textcolor{black}{to which we add terms relevant in describing interactions with} external magnetic fields, via the Zeeman effect.
\textcolor{black}{With this representation of the Hamiltonian for} the atomic system, we proceed to calculate the \textcolor{black}{eigenvalues, which correspond to the} transition energies, and absolute line strengths \textcolor{black}{from the dipole matrix elements}.
\textcolor{black}{As a final step, we incorporate the Doppler effect in our model via} a Voigt profile (convolution of the Lorentzian homogeneous linewidth and the Gaussian from the Maxwellian velocity distribution) \textcolor{black}{for} each of the atomic transitions as defined by their energy and strength.
\textcolor{black}{In this manner, the total susceptibility of the medium is calculated as the sum of the individual susceptibilities for all dipole-allowed transitions, each with an effective detuning in the scope of a global linear detuning}.
\textcolor{black}{At this point it is convenient to split the susceptibility into separate components for $\sigma^\pm$ and $\pi$ transitions}.

\begin{figure}[t]
\includegraphics[width=0.5\columnwidth]{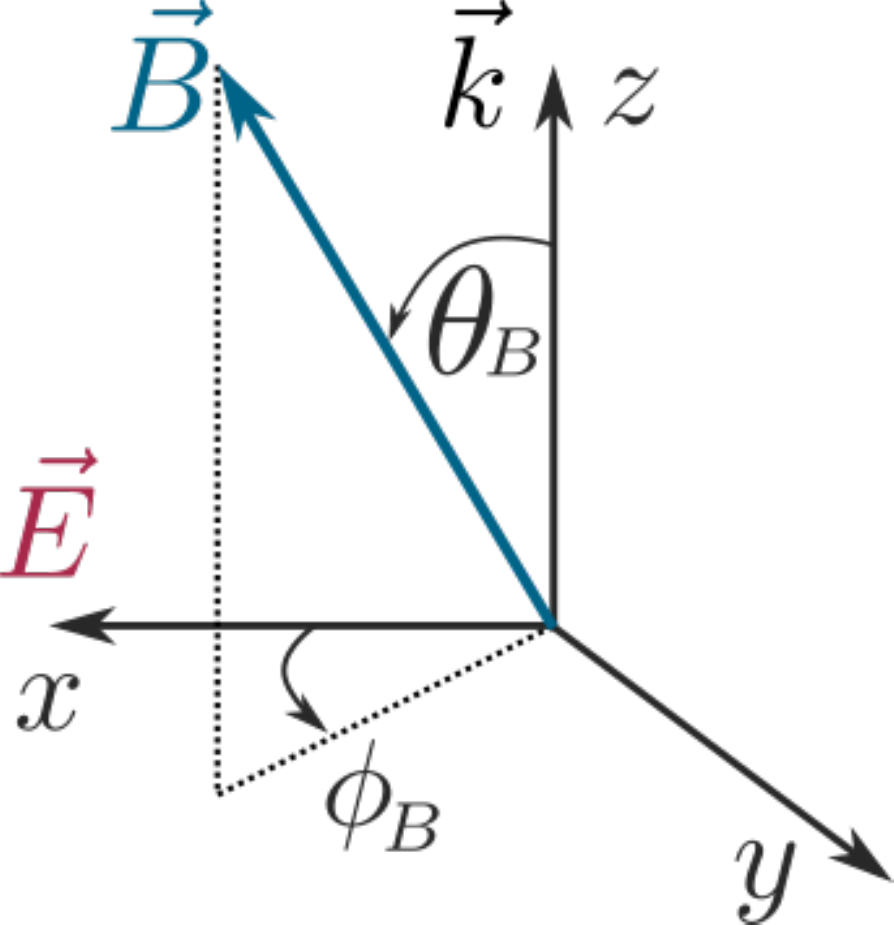}
\caption{General geometry for experiments involving an externally applied magnetic field and light; for the Voigt geometry where $\vec{k}\perp \vec{B}$, $\theta_{B}=\pi/2$. The light is assumed to be linearly polarised \textcolor{black}{in the $x-y$ plane, with the electric field vector $\vec{E}$ taken here} to be vertically polarised along the $x$-axis. The external magnetic field \textcolor{black}{$\vec{B}$} makes an angle $\phi_{B}$ with respect to $\vec{E}$, which \textcolor{black}{when projected onto the $x-y$ plane gives components} parallel and perpendicular \textcolor{black}{to $\vec{E}$. The former drives $\pi$ transitions and the latter $\sigma^{\pm}$ transitions.}}
\label{fig:geometry}
\end{figure}

For a given total susceptibility the propagation of light through the atomic medium is done by solving the corresponding wave equation and finding the two propagation eigenmodes.
Each eigenmode \textcolor{black}{is expressed in terms of the components of the susceptibility, which allow the light to couple to the transitions in distinct ways.
One can also find a relation between the eigenmodes and the complex refractive indices of the medium.}
The exact coupling \textcolor{black}{of the eigenmodes to the transitions} depends on the geometry of the \textcolor{black}{atom-light} system.
In \textcolor{black}{our experiment} the system is set up in the Voigt geometry \textcolor{black}{as shown in figure~\ref{fig:geometry}, which consists of setting} the externally applied magnetic field vector $\vec{B}$ \textcolor{black}{perpendicular to the light wavevector $\vec{k}$.}
\textcolor{black}{We further constrain the geometry of the system by taking} the electric field vector of \textcolor{black}{our laser beam} $\vec{E}$ along the Cartesian $x$-axis, which \textcolor{black}{allows us to consider our beam} as a linearly polarised plane-wave propagating along the $z$-axis and polarised in the $x-y$ plane.
In this manner, for the Voigt geometry we take the external magnetic field vector in the $x-y$ plane, with an angle $\phi_{B}$ defined as the direction of said vector relative to the $x$-axis ($\vec{E}$).
\textcolor{black}{This angle also gives a simple way of describing} the relative coupling of the light to the atomic electron transitions as a consequence of the projections of $\vec{B}$ onto $\vec{E}$, which in turn change the angular-momentum algebra.
For a projection of $\vec{B}$ parallel to the polarisation of the light (\textit{i.e.} $\vec{E}$) $\pi$ transitions with $\Delta m_{J}=0$ are driven, while for a projection of $\vec{B}$ perpendicular to the polarisation of the light the $\sigma^{\pm}$ transitions, with $\Delta m_{J}=\pm1$, are driven~\cite{f2f}.
\textcolor{black}{Circular polarisation of left- or right-handedness in the Voigt geometry can be treated by assuming a linear combination of the $x$ and $y$ components of linearly polarised light as there is no immediate effect of the relative phase between the two.
In this particular case, the relative phase $\phi=(2n-1)\pi/4$, with $n$ an integer, is merely reflected in the strength of each component, which is equal for the $x$ and $y$ components.}
\textcolor{black}{A corollary of the above is that the system exhibits equivalent solutions for values of $\phi_{B}$ modulo $\pi$, and this in turn results in the inability to determine the absolute direction of the magnetic field with just one measurement.}

As mentioned \textcolor{black}{above, the propagation through the atomic medium of each of the eigenmodes has an associated complex refractive index.}
In general\textcolor{black}{, the medium has two unique refractive indices, which in light of them being complex, gives rise to dichroism and birefringence}.
To take these effects into account\textcolor{black}{, the electric field of the light is propagated through the atomic medium.
The input electric field is first transformed into the coordinate system of the medium's eigenbasis, which corresponds to the Cartesian basis in the Voigt geometry, with the orthogonal components coinciding with the parallel and perpendicular directions to the external magnetic field vector.
Once the field has been written in this new basis it is propagated using each index, $n_{i}$, over a distance $L$ in the medium by using the multiplicative factor $e^{{\rm{i}}n_{i} k L}$.
The resulting electric field is then transformed back into the laboratory coordinate system.
At this stage the output electric field can be conveniently analysed }
via Stokes polarimetry~\cite{Schaefer2007, Weller2012}.
\textcolor{black}{This formalism provides a set of parameters which are easily obtained with standard laboratory equipment and measurements of the intensity of light in different sets of orthogonal polarisation bases.}
The Stokes parameter $S_0$ \textcolor{black}{is independent of the measurement basis, and for an incident beam of intensity $I_{0}$} is defined as
\begin{eqnarray}
	S_{0} \equiv \frac{I_{x} + I_{y}}{I_{0}} = \frac{I_{\nearrow} + I_{\searrow}}{I_{0}} = \frac{I_{\rm{RCP}} +  I_{\rm{LCP}}}{I_{0}}.
	\label{eqn:S0}
\end{eqnarray}
\textcolor{black}{Physically, $S_0$ represents the normalised total transmitted intensity.}
The remaining Stokes parameters, $S_{1},\,S_{2}\,\&\, S_{3}$,
\begin{eqnarray}
	S_{1} \equiv& \displaystyle\frac{I_{x} - I_{y}}{I_{0}}, \label{eqn:S1}\\
	S_{2} \equiv& \displaystyle\frac{I_{\nearrow} - I_{\searrow}}{I_{0}}, \label{eqn:S2}\\
	S_{3} \equiv& \displaystyle\frac{I_{\rm{RCP}} - I_{\rm{LCP}}}{I_{0}}, \label{eqn:S3}
\end{eqnarray}
give information on the optical rotation of light generated by the atomic medium by looking at the differences between the orthogonal polarisation components in the horizontal ($I_{x},I_{y}$), diagonal ($I_{\nearrow},I_{\searrow}$) and circular bases ($I_{\rm{RCP}},I_{\rm{LCP}}$), respectively.

\begin{figure}[t]
\includegraphics[width=\columnwidth,clip=true,trim=0mm 0mm 0mm 0mm]{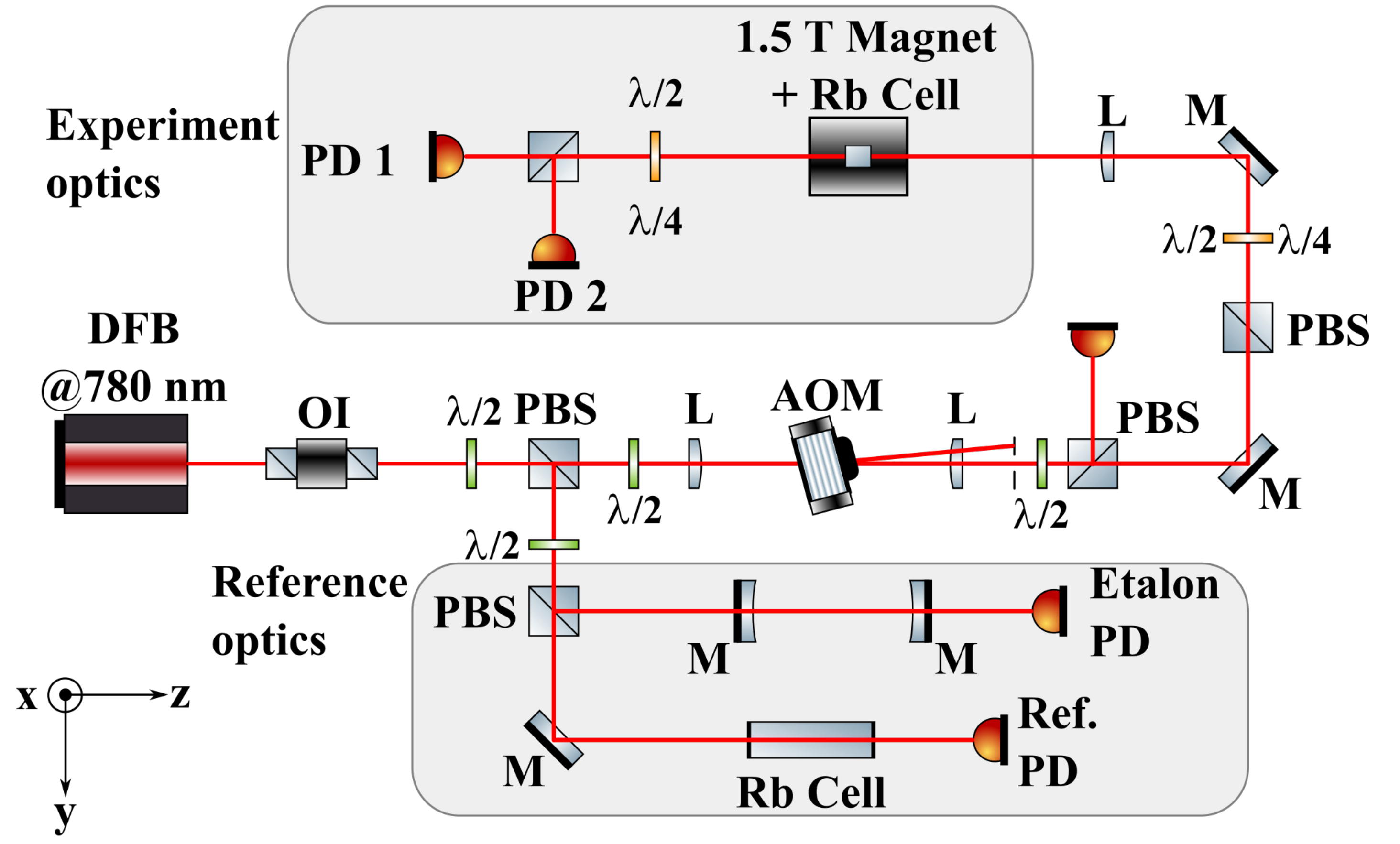}
\caption{\textcolor{black}{Schematic of the most relevant components of the optical setup used for acquiring experimental data. Light from a distributed feedback (DFB) laser passes through an optical isolator (OI) and is split into two components. Approximately half of the light is used for a reference optical system used to calibrate the laser scan in a zero magnetic field environment; the system consists of a room temperature 75~mm natural abundance Rb reference cell and a Fabry-P\'erot etalon made with two mirrors (M)}. The rest of the light is sent to the experiment cell via an acousto-optic modulator (AOM) that can be used to keep the power constant~\cite{Truong2012} and polarising beamsplitter (PBS) cubes and half-/quarter-wave plates ($\lambda/2\, \& \, \lambda/4$, respectively) to ensure the input light is in the desired polarisation state. \textcolor{black}{The 1~mm long experiment cell contains isotopically enriched $^{87}$Rb ($99\%$ purity)}, and is placed in a cylindrical magnet designed to give a primarily axial field of 1.62~T at its centre~\cite{Trenec2011}. A polarisation-sensitive detection scheme is used after the experiment cell, consisting of a PBS, retarder wave plate and two balanced photodiodes (PD), that allow a voltage signal with information on the optical rotation and absorption of the medium to be measured. Plano-convex lenses (L) are used to resize the beam at different points in order to avoid significant clipping.}
\label{fig:setup}
\end{figure}

\begin{figure}[tb]
\includegraphics[width=0.9\columnwidth,clip=true,trim=0mm 0mm 0mm 0mm]{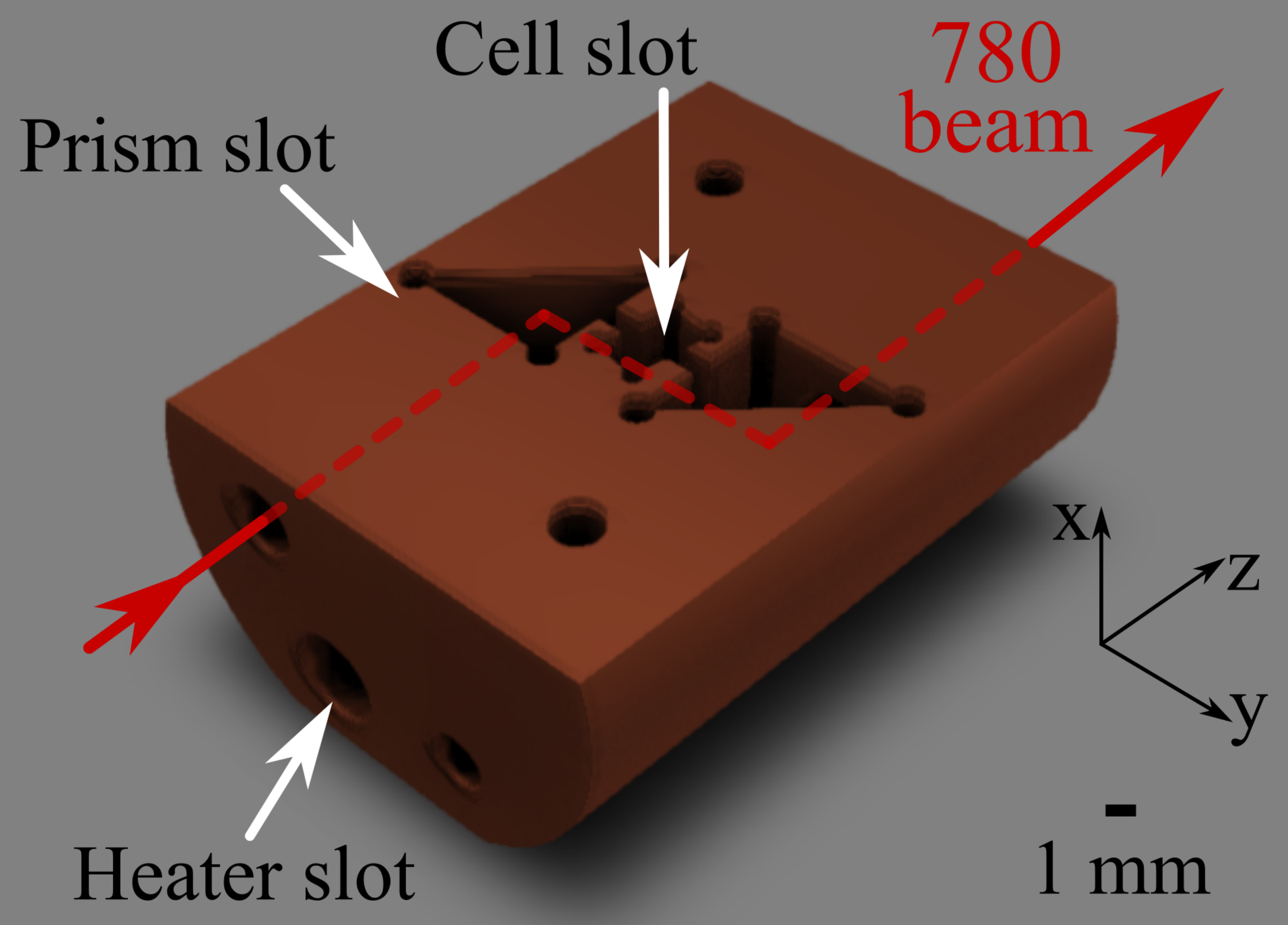}
\caption{Internal copper heater block used in the experiments in the Voigt geometry. Shown here is the main copper block, with an angled slot for the $1$~mm cell to avoid back-reflection from the cell windows and two spaces for $5$~mm right-angle prisms. Light enters and exits the heater block through two optical access holes (seen in the upper-left) and passes through the cell perpendicular to its original direction of incidence. The larger hole (bottom-centre) houses a resistive heater element that serves to heat the entire block in order to raise the temperature of the vapour in the interior of the cell to the desired point. This block is covered by a copper lid and housed in a custom PTFE cylinder that preserves the optical access and cable feed-throughs necessary to carry out experiments when the whole assembly is located inside the permanent magnet.}
\label{fig:heater}
\end{figure}

\begin{figure*}[tb]
\includegraphics[width=2.075\columnwidth,clip=true,trim=1mm 2mm 1mm 1mm]{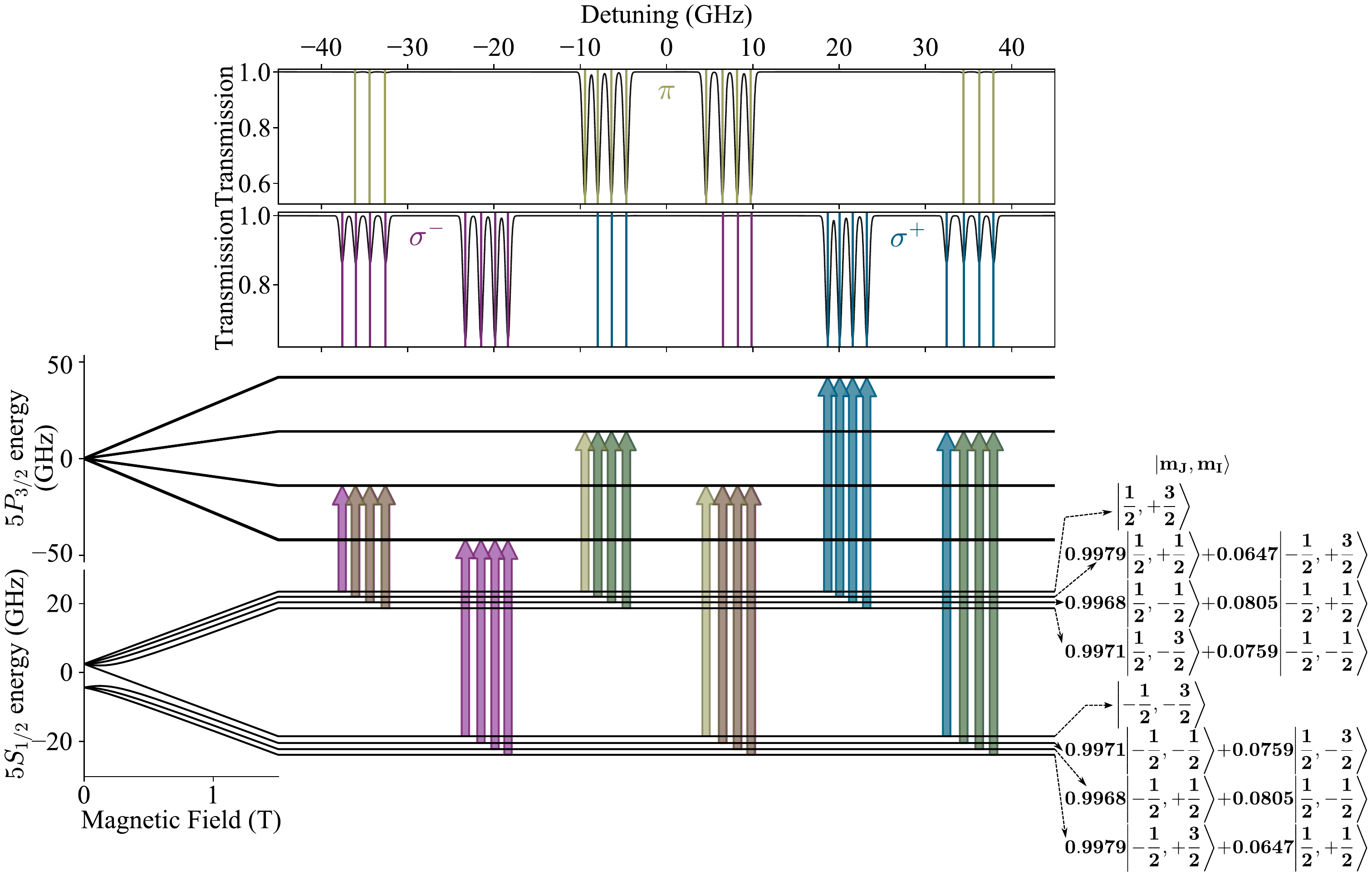}
\caption{Spectroscopy of the Rb D2 line \textcolor{black}{for a system with an external magnetic field of 1.5~T arranged in the Voigt geometry}. The bottom-left \textcolor{black}{of the} figure shows the evolution of the \textcolor{black}{ground and excited state energy levels in $^{87}$Rb, corresponding to the} 5$S_{1/2}$ and 5$P_{3/2}$ atomic energy levels, \textcolor{black}{respectively, with an increasing} magnetic field strength of up to 1.5~T. \textcolor{black}{In the lower right of the figure,} the ground-state levels involved in the atomic transitions at 1.5~T, including the admixtures due to remnant effects of the hyperfine interaction, \textcolor{black}{are given in the $|m_{J},m_{I}\rangle$ basis}. The upper panels show the calculated spectra, \textcolor{black}{as functions of linear detuning,} for $\pi$ (olive) transitions ($\vec{E} \parallel \vec{B}$), and $\sigma^+$ (blue), $\sigma^-$ (purple) transitions ($\vec{E} \perp \vec{B}$). The coloured arrows \textcolor{black}{in the energy level diagram point from the initial to the final} states involved in the transition, with overlapping arrows showing the presence of weaker transitions due to the admixtures present in the system.}
\label{fig:bigdiagram}
\end{figure*}

\section{Experimental setup}
\label{sec:expt}

\textcolor{black}{Experimental work for this paper was carried out using the optical setup described in figure~\ref{fig:setup}}.
\textcolor{black}{The laser used is a tuneable distributed feedback (DFB) laser, with a central wavelength of 780~nm and }a quoted linewidth of $<2$~MHz.
The frequency is tuned by changing the temperature of the laser chip, which allows a frequency mode-hop free tuning range over many hundreds of GHz.
A typical scan of this size takes $\sim 2$~seconds and is limited by the response time of the temperature control circuit of the laser.
\textcolor{black}{The emitted light from the laser is split using polarisation optics after an optical isolator to maintain high polarisation purity and manage the beam intensity.
Some of the light is taken to a secondary optical setup for calibration of the laser frequency scan.
A signal from a 75~mm natural abundance Rb reference cell provides an absolute frequency reference, whereas a transmission signal from a Fabry-P\'erot etalon is used to linearise the scan}.
\textcolor{black}{These two signals are then processed as} outlined in ref.~\cite{Keaveney2014a} to obtain spectra as a function of linear detuning.
The remainder of the light is then passed through several polarising elements that allow us to obtain a well-defined linear polarisation to use in our experiment cell.
A circular polarisation can be obtained by replacing the half-wave retarder plate ($\lambda/2$) for a quarter-wave retarder plate ($\lambda/4$) before the light goes through the experiment cell.

The experiments are carried out in a \textcolor{black}{cuboidal vapour cell with a cavity length of 1~mm containing isotopically enriched $99\%$ purity $^{87}$Rb.}
\textcolor{black}{In order to provide a large enough atomic density and optical depth for observing the desired phenomena the cell must be heated.}
Microfabricated cells, such as this one, have previously been found to exhibit birefringent properties when heated to their optimal operating temperatures \cite{Weller2013,Sargsyan2006,Jahier2000,Steffen2013}.
Details on the fabrication of this cell can be found in \cite{Knappe2005}.
The cell and beam-steering optics (uncoated right-angle prisms) are \textcolor{black}{placed in a custom-made copper holder} which also houses an internal heater, as seen in figure~\ref{fig:heater}.
During operation the temperature is set by applying a voltage to the heater to raise the temperature of the copper surrounding the cell.
\textcolor{black}{The stability of the holder is maintained} passively by fixing the voltage and allowing \textcolor{black}{thermalisation of the metal block} with the surroundings.
The copper bed is \textcolor{black}{enclosed in a secondary cylinder made of PTFE that acts as a shield and} is only in weak thermal contact with it in order to ensure there are no sharp fluctuations in the temperature of the system.
Due to the reduced footprint of the system imposed by the central bore size of the magnet it was not possible to include a suitable temperature sensor in this mount.

The external magnetic field is generated by a cylindrical permanent magnet, with a central bore of diameter $22$~mm along its axis, designed using the ``magic sphere'' configuration described in \cite{Trenec2011}.
The maximum value of the field produced by the magnet is 1.62~T at its centre, with the field strength quickly falling radially outwards to the ends of the magnet.
The cylindrical PTFE assembly housing the copper heater block with the experiment cell sits inside the magnet's bore.
This particular design allows for a well-characterised field along the axis of the magnet, which in turn ensures field homogeneity across the length of the experiment cell, while maintaining the field on the outside of the magnet to below hundreds of mT.
Further details on the construction and characterisation of the magnet can be found in ref.~\cite{Trenec2011}.

\textcolor{black}{The experiment is conducted in the weak-probe regime~\cite{Sherlock2009} to avoid optical pumping.
Therefore the optical power of the interrogating laser beam (\textit{i.e.} the probe beam) is maintained below the saturation intensity.
For a beam waist ($1/e^{2}$) of approximately $0.7$~mm this means keeping the power close to } 1~$\rm{\mu}$W.
\textcolor{black}{A probe beam of these characteristics gives an} effective spatial resolution \textcolor{black}{roughly equal to the }volume of atoms interrogated by the laser beam, which \textcolor{black}{can be considered to be a cylinder whose length is that of the experiment vapour cell ($1$~mm) and whose radius is the beam waist.}
It is worth nothing that the length of the cylindrical heater block assembly previously described ($28$~mm) is smaller than the length of the magnet ($152$~mm) and this generates difficulties in aligning the heater block axis relative to the laboratory frame of reference once inside the magnet.
As such, the heater block is slightly rolled about the axis of the magnet ($z$) which in turn results in a relative orientation of the $x,y$ axes of the atom and the laboratory $x,y$ axes that can be described by an effective offset in the angle $\phi_{B}$. 

\section{Analysis of atomic absorption spectra at high fields}
\label{sec:spectra}

Figure~\ref{fig:bigdiagram} is created using \textcolor{black}{the theoretical model described in section \ref{sec:theory} and shows the evolution of the} atomic energy levels as a function of magnetic field strength.
\textcolor{black}{In addition,  the predicted absorption spectra in a 1.5~T external magnetic field are depicted}.
\textcolor{black}{In the lower panel the ground- and excited-state energy level diagrams show, respectively, the states in the 5$S_{1/2}$ and 5$P_{3/2}$ manifolds at 1.5~T as well as the initial and final states involved in each of the allowed transitions}.
\textcolor{black}{Seen in the top two panels are the }calculated absorption for an isotopically enriched \textcolor{black}{vapour cell ($99\%$ purity $^{87}$Rb)} at $100^{\circ}$C, \textcolor{black}{separated into} $\pi$ (upper panel) and $\sigma^\pm$ (lower panel) transitions.
\textcolor{black}{The position of the vertical lines serves to indicate the atomic resonances in terms of} the linear frequency.
\textcolor{black}{The colour of these lines corresponds to different types of transitions}: $\pi$ transitions are olive green, $\sigma^{+}$ transitions are blue and $\sigma^{-}$ transitions are purple.
\textcolor{black}{For the ground state, the decompositions} in the $m_{J},m_{I}$ basis are \textcolor{black}{given to the right of the corresponding sublevel}.
The arrows, with colours again corresponding to the type of transition as previously mentioned, are semi-transparent to evidence the fact there are still some overlapping transitions due to a small remnant admixture in the state decomposition due to hyperfine interactions.
This can also be seen in the respective detunings of the transitions on the top two panels, where $\pi$ and $\sigma^{\pm}$ have frequencies that coincide between them.

\begin{figure}[tb]
\includegraphics[width=1.0\columnwidth,clip=true,trim=2mm 2mm 0mm 0mm]{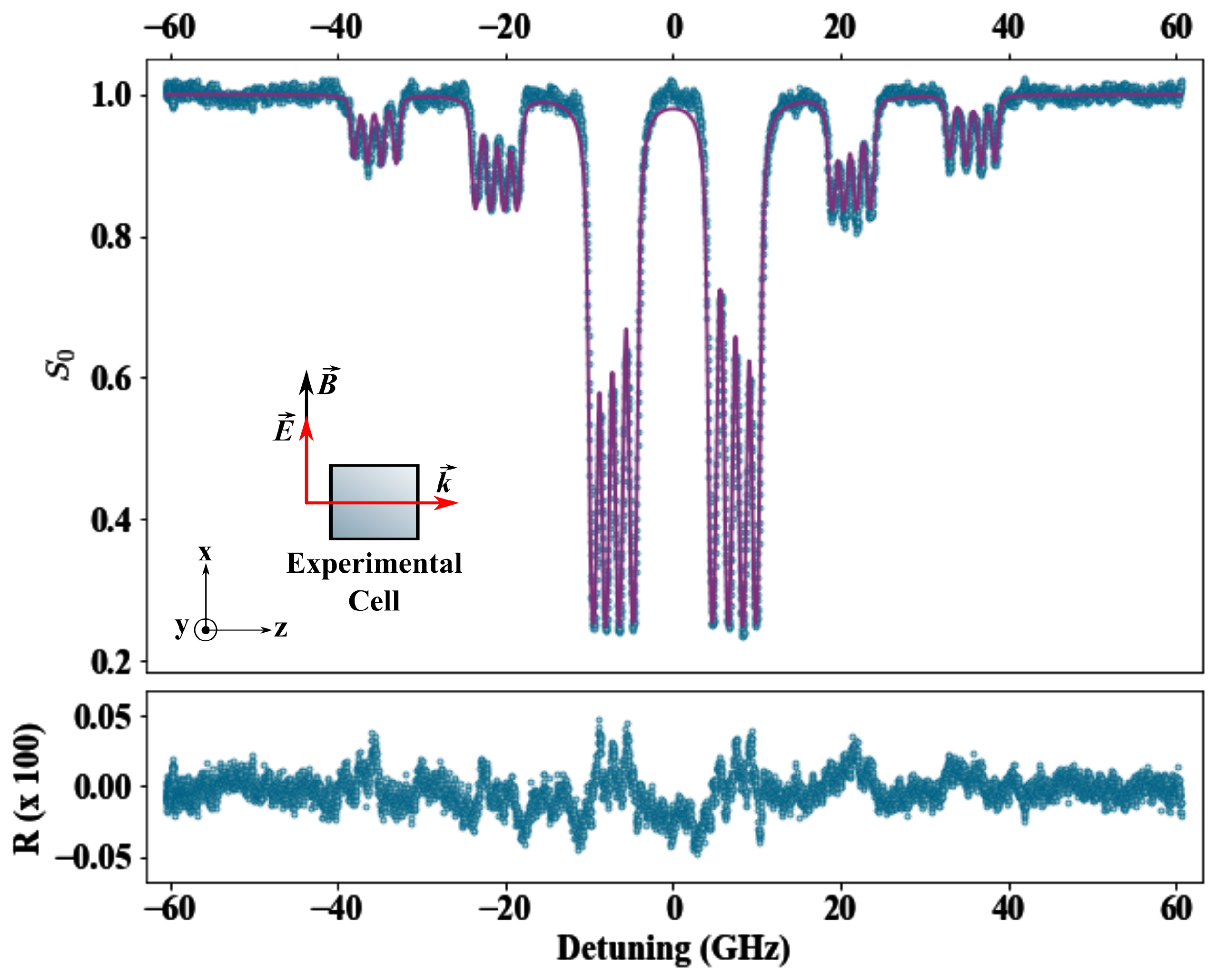}
\caption{Experimental data (blue circles; colour online) \textcolor{black}{expressed in terms of the linear detuning of the laser scan} for a horizontally polarised input beam with the corresponding fit (purple line; colour online) using our {\it ElecSus} model, with residuals shown (bottom panel). \textcolor{black}{Data and the theoretical fit are in very good agreement} (RMS error of 1.2\%). For this spectrum the free parameters in the fit are: $\phi_{\rm{B}}$, the angle of the magnetic field with respect to the $x$-axis, $B$, the magnetic field strength and $T$, the temperature of the atoms. Average values of $\phi_{\rm{B}} = (0.4491\pm0.0007)$~rad, $B = (1.52 \pm 0.08)$~T and $T = (108.94 \pm 0.04)^{\circ}$C are obtained from fitting five spectra. All other parameters for the system are fixed as follows: $\theta_{\rm{B}}=\pi / 2$, $\Gamma_{\rm{Buff}}=350$~MHz and $\delta_{\rm{shift}}=50$~MHz.}
\label{fig:Hpol_fit}
\end{figure}

\begin{figure}[tb]
\includegraphics[width=1.0\columnwidth,clip=true,trim=2mm 2mm 0mm 0mm]{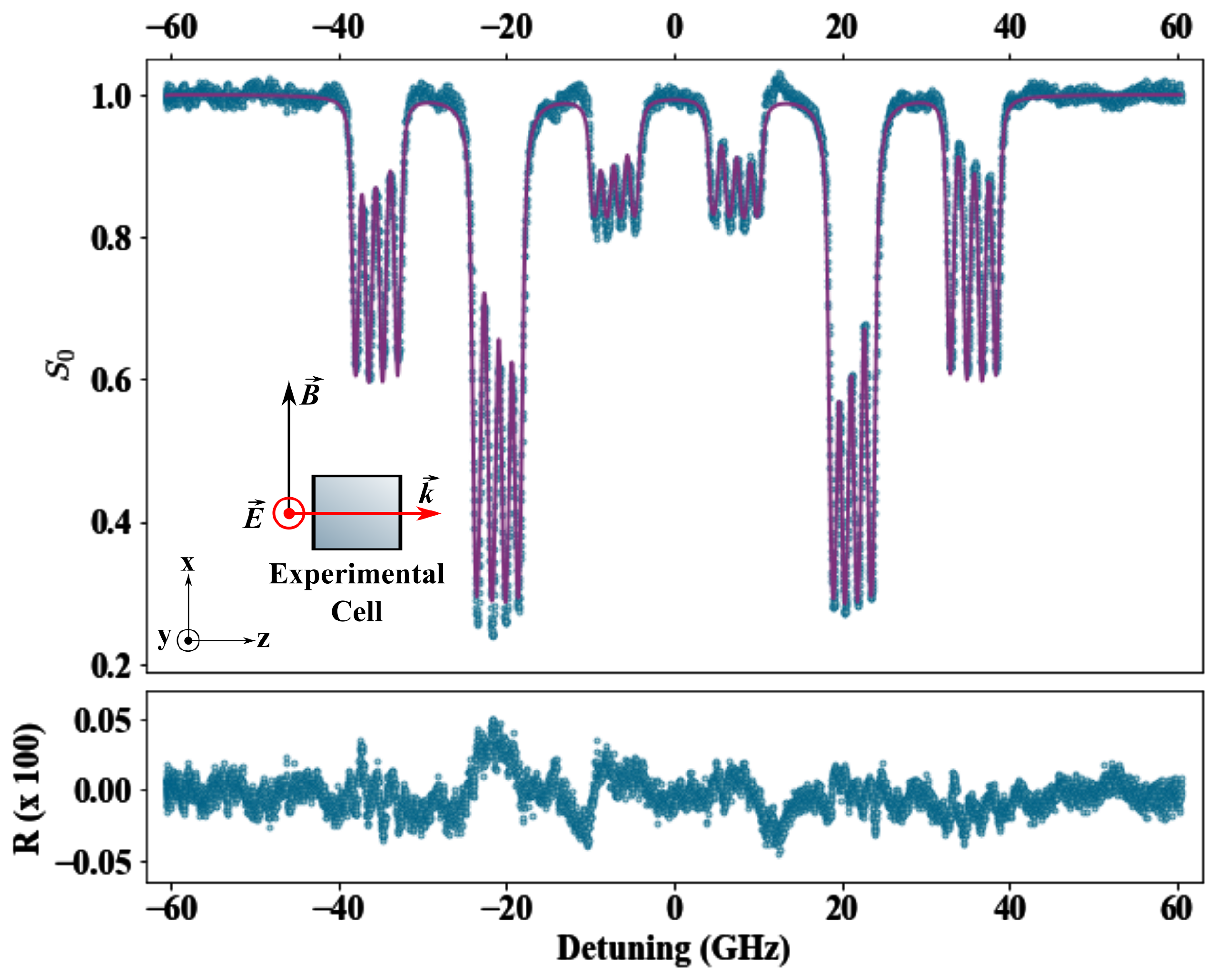}
\caption{Experimental data (blue circles; colour online) \textcolor{black}{expressed in terms of the linear detuning of the laser scan} for a vertically polarised input beam with the corresponding fit (purple line; coloured online) using our {\it ElecSus} model, with residuals shown (bottom panel). \textcolor{black}{Data and the theoretical fit are in very good agreement} (RMS error of 1.2\%). For this spectrum the free parameters in the fit are: $\phi_{\rm{B}}$, the angle of the magnetic field with respect to the $x$-axis, $B$, the magnetic field strength and $T$, the temperature of the atoms. Average values of $\phi_{\rm{B}} = (2.0082\pm0.0007)$~rad, $B = (1.52 \pm 0.07)$~T and $T = (110.23 \pm 0.03)$C are obtained from fitting five spectra. All other parameters for the system are fixed as follows: $\theta_{\rm{B}}=\pi / 2$, $\Gamma_{\rm{Buff}}=350$~MHz and $\delta_{\rm{shift}}=50$~MHz.}
\label{fig:Vpol_fit}
\end{figure}

\textcolor{black}{At a field strength of 1.5~T,} the 5$P_{3/2}$ states strongly decouple into the $m_{J}, m_{I}$ basis \textcolor{black}{and result in four multiplets organised by} $m_{J} = 3/2,1/2,-1/2,-3/2$ projections.
Furthermore, at this magnetic field strength the ground state has a much more complete decoupling than that observed at lower field strengths \cite{Keaveney2019,Zentile2014a,Whiting2015,Whiting2016a,Whiting2017,Whiting2017b}.
As such, we can treat the system as being completely in the Hyperfine Paschen-Back (HPB) regime.
This means that the initial hyperfine ground states are now split into two distinct groups, corresponding to the projections $m_{J} = \pm1/2$, with the $m_{I} = 3/2,1/2,-1/2,-3/2$ states clearly defined despite the Doppler-broadening in the vapour.
This gives us well-defined multiplets of four `strong' transitions ($\vert m_J, m_I\rangle \rightarrow \vert m_{J}', m_{I}\rangle$, with $m_{J}' = m_{J} \; (\pi), m_{J}+1 \; (\sigma^+), m_{J}-1 \; (\sigma^-)$).
It is worth adding that at this point there are still some `weak' transitions present \textcolor{black}{due to} the ground states not being pure eigenstates in the $m_{J},m_{I}$ basis.
\textcolor{black}{This is evident in the small ($<1\%$)} admixture of the opposite $m_{J}$ state in the state decomposition on the bottom right of fig.~\ref{fig:bigdiagram} (more details can be found in ref.~\cite{Zentile2014a}).

Figures~\ref{fig:Hpol_fit} and \ref{fig:Vpol_fit} show experimental data, averaged over five spectra, that have been fitted using \textcolor{black}{the theory obtained with \textit{ElecSus}.}
\textcolor{black}{From the fit, the RMS error between data and theory is calculated for} figure~\ref{fig:Hpol_fit}(\ref{fig:Vpol_fit}) \textcolor{black}{to be} 1.2\%.
\textcolor{black}{The residuals R, multiplied by a factor of 100 for ease of viewing, are also shown.
The absence of obvious  structure in the residuals, and the large multiplier required for the residuals, are indicators of a very good fit \cite{Hughes2010}}.
The fit is carried out with three free parameters: $\phi_{\rm{B}}$, the angle between the magnetic field and the direction of polarisation of the laser beam (see figure~\ref{fig:geometry}), taken to be linear along the $x$($y$)-axis; $B$, the magnitude of the magnetic field the atoms are exposed to and $T$, the temperature of the atoms in the experiment cell.
Other significant experimental parameters, such as those relating to the effects due to buffer gas in the cell, such as the amount of inhomogeneous broadening $\Gamma_{\rm{Buff}}$ caused by collisions and a shift in the frequency of the transitions $\delta_{\rm{shift}}$, are kept fixed.
The values for these fixed parameters are obtained \textit{a priori} by fitting other spectra similar to those averaged and shown in figures~\ref{fig:Hpol_fit} and \ref{fig:Vpol_fit}.
Of the remaining parameters in the fit, the field angle $\theta_{\rm{B}}$ is fixed by the geometry of the experimental setup in the Voigt geometry ($\theta_{\rm{B}} = \pi/2$). 
We attribute the significant buffer gas broadening in the spectra to He atoms trapped in the cell, after the cell was exposed to a He environment in previous experiments and note \textcolor{black}{that this does not generate important shifts} in any of the resonance lines.
Using the literature values of the broadening coefficient for He~\cite{Rotondaro1997}, we extract a pressure of $\sim 18$~torr ($\sim 24$~mbar) for the amount of said buffer gas in our experiment cell.
Note that while the time necessary to acquire a spectrum is on the order of a second, the time needed to analyse the data and generate a fit is on the order of minutes due to the complexity of the parameter space.

For the spectrum shown in figure~\ref{fig:Hpol_fit} we obtain a value of $\phi_{\rm{B}} = (0.4491 \pm 0.0007)$ radians ($(25.74\pm0.04)^{\circ}$).
Similarly, \textcolor{black}{for} the fit shown in figure~\ref{fig:Vpol_fit} we obtain a value of $\phi_{\rm{B}} = (2.0081 \pm 0.0007)$ radians ($(115.05\pm0.04)^{\circ}$).
Both of these values differ from their corresponding expected values by $\approx 0.45$ radians ($\sim 25^{\circ}$), which we take as a systematic error due to the orientation of the cell heater block inside the bore of the cylindrical magnet used in the experiment.
As a result, there is excitation of both $\pi$ and $\sigma^{\pm}$ transitions in both spectra shown due to the presence of parallel and perpendicular components of $\vec{B}$ along the direction of polarisation of the light.
In this case, the difference in strength between the transitions is given as a simple factor of $\cos^{2}(\phi_{\rm{B}})$ for the parallel component and $\sin^{2}(\phi_{\rm{B}})$ for the perpendicular component.
According to the difference in the expected value of $\phi_{\rm{B}}$ obtained from the fits, this results in an approximate $4:1$ ratio in the strengths of the lines; this is clearly visible in both figure~\ref{fig:Hpol_fit} and \ref{fig:Vpol_fit}.

Similarly, for both spectra shown in figures~\ref{fig:Hpol_fit} and \ref{fig:Vpol_fit} we obtain a value $B = 1.52$~T.
The uncertainties in these values, $80$~mT and $70$~mT, respectively for the two spectra, can be mainly attributed to linearity of the laser scan in our experiment.
The DFB laser used in our experiment allows for a large mode-hop-free scan ($\sim 150$~GHz) at the expense of a non-linearity that is introduced as the frequency is changed.
\textcolor{black}{It is this non-linearity, along with other systematic errors in the linearisation and calibration of} the frequency axis, \textcolor{black}{that }is the primary source of uncertainty in our measurements.
In future work we plan to design the experiment so that this non-linearity can be reduced in order to improve the precision in our measurement of the magnetic field strength.

Furthering the ideas and work presented in \cite{Keaveney2019}, we propose this as an atomic technique for measuring large magnetic fields and their relative orientation.
With the system completely in the hyperfine Paschen-Back regime, the Zeeman shift presented in all the resonance line positions allows for a better determination of the magnetic field strength.
In addition to this, the relative strength between sets of transitions, due to different coupling strengths, is better observed.
\textcolor{black}{This in turn} allows a more precise determination of the relative \textcolor{black}{orientation of the direction of polarisation (\textit{i.e.} of the electric field vector) of the light with respect to the external magnetic field}.
Thus, the present system and technique lead to a natural application of atomic-based spectroscopy in vector magnetometry.

\section{Sensitivity of optical rotation signals to the birefringence of cell windows}
\label{sec:birref}

\begin{figure}[tb]
\includegraphics[width=1.0\columnwidth,clip=true,trim=0mm 2mm 0mm 0mm]{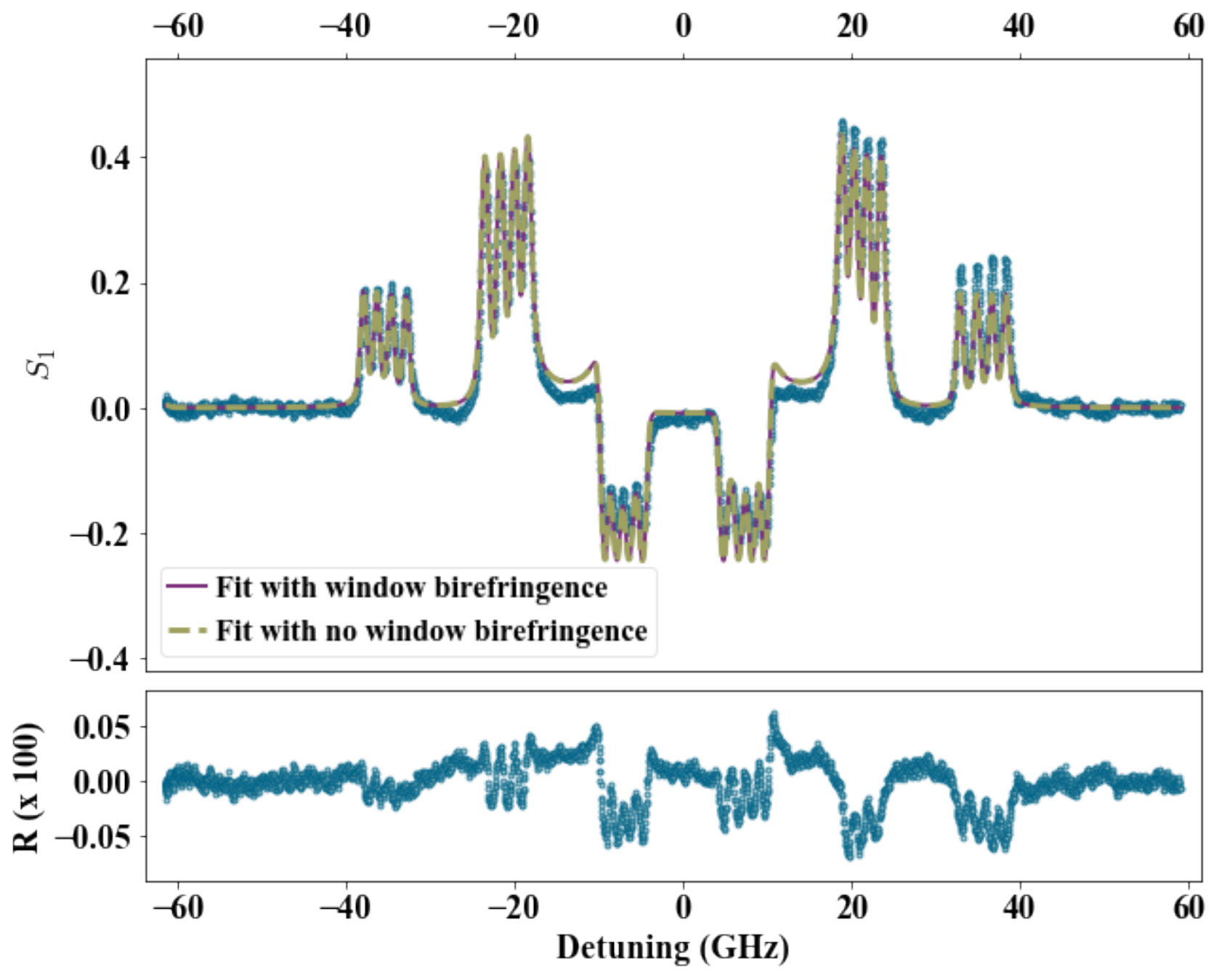}
\caption{Experimental data (blue circles) taken as a function of linear detuning for a linearly polarised input beam with the corresponding fit (purple line) of the $S_{1}$ parameter taking window birefringence into account, with residuals R shown. \textcolor{black}{Data and the theoretical fit are in very good agreement} (RMS error of $\sim$2\%). In this case the fixed parameters are $\Gamma_{\rm{Buff}}=350$~MHz, $\delta_{\rm{shift}}=50$~MHz and $\theta_{\rm{B}} = \pi / 2$, while the fit allows $T,B,\phi_{\rm{B}}$ to float; also included in the fit are the parameters to take into account the birefringence effects of the cell windows ($\theta_{\rm{BR}},\phi_{\rm{BR}}$). A fit without the effect of the cell window birefringence (broken line) is included for comparison.}
\label{fig:S1_fit}
\end{figure}

\begin{figure}[tb]
\includegraphics[width=1.0\columnwidth,clip=true,trim=0mm 2mm 0mm 0mm]{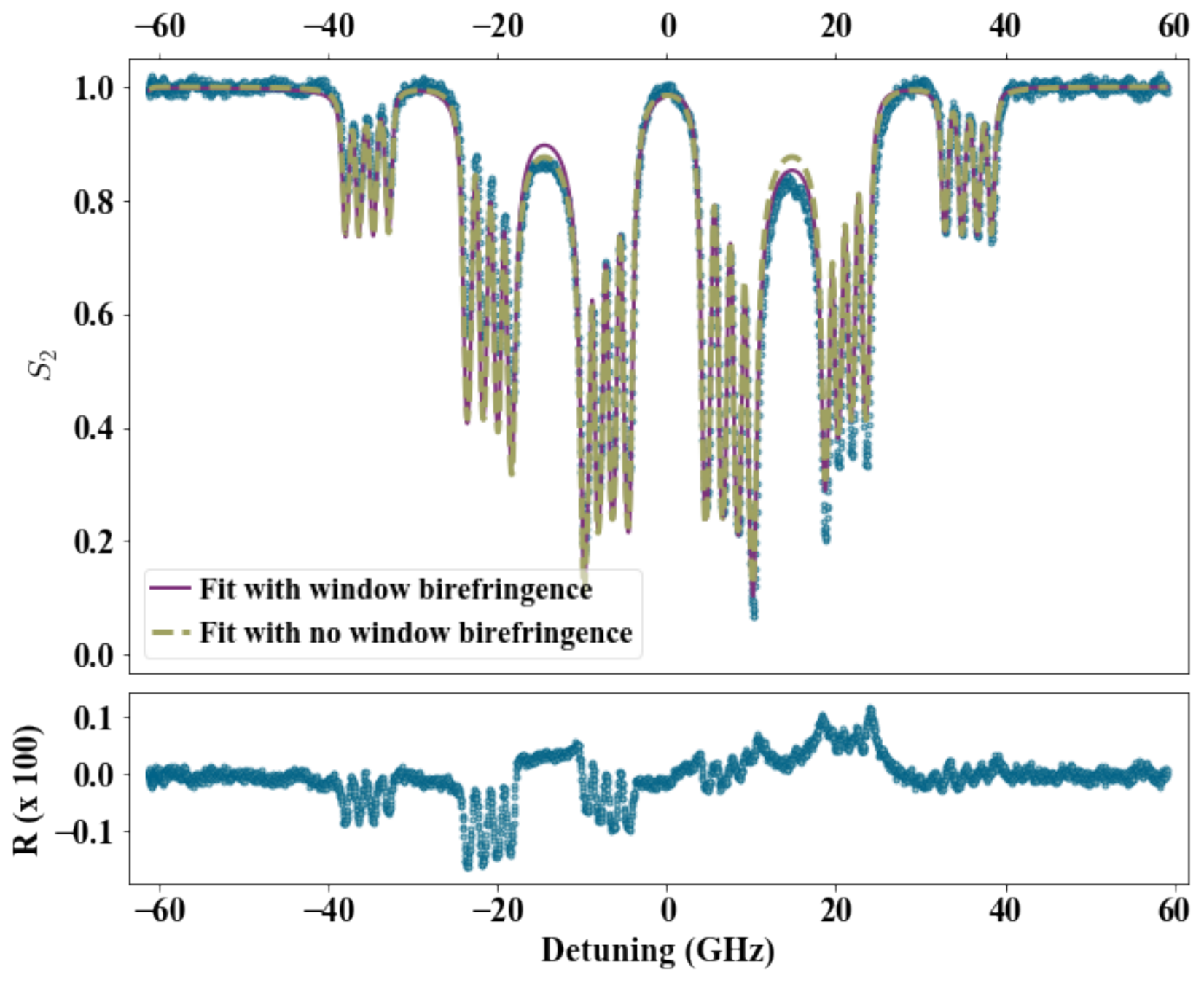}
\caption{Experimental data (blue circles) taken as a function of linear detuning for a linearly polarised input beam with the corresponding fit (purple line) of the $S_{2}$ parameter taking window birefringence into account, with residuals R shown. \textcolor{black}{Data and the theoretical fit are in very good agreement} (RMS error of $\sim$3\%). In this case the fixed parameters are $\Gamma_{\rm{Buff}}=350$~MHz, $\delta_{\rm{shift}}=50$~MHz and $\theta_{\rm{B}} = \pi / 2$, while the fit allows $T,B,\phi_{\rm{B}}$ to float; also included in the fit are the parameters to take into account the birefringence effects of the cell windows ($\theta_{\rm{BR}},\phi_{\rm{BR}}$). A fit without the effect of the cell window birefringence (broken line) is included for comparison.}
\label{fig:S2_fit}
\end{figure}

\begin{figure}[tb]
\includegraphics[width=1.0\columnwidth,clip=true,trim=0mm 2mm 0mm 0mm]{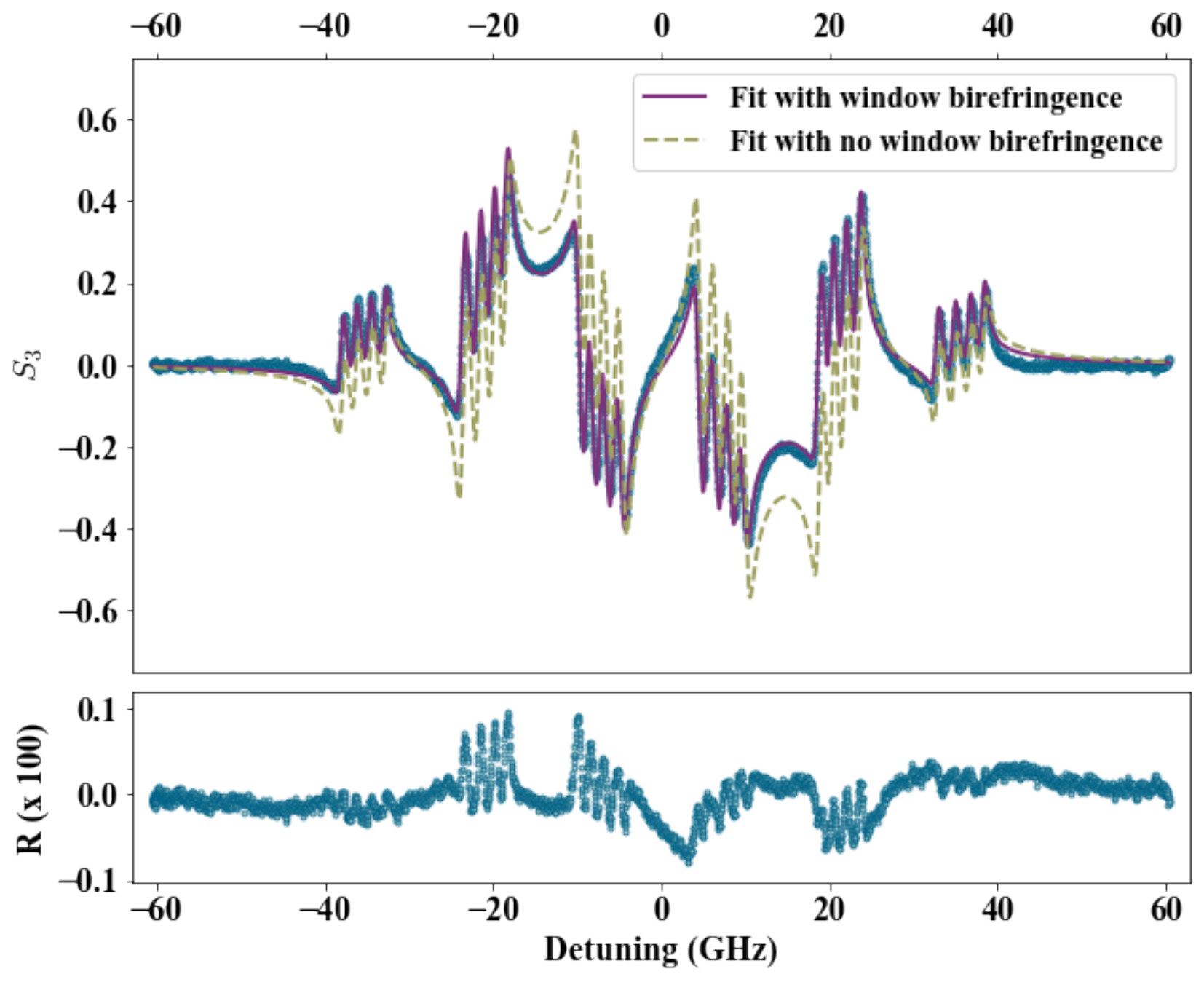}
\caption{Experimental data (blue circles) taken as a function of linear detuning for a linearly polarised input beam with the corresponding fit (purple line) of the $S_{3}$ parameter taking window birefringence into account, with residuals R shown. \textcolor{black}{Data and the theoretical fit are in very good agreement} (RMS error of $\sim$2\%). In this case the fixed parameters are $\Gamma_{\rm{Buff}}=350$~MHz, $\delta_{\rm{shift}}=50$~MHz and $\theta_{\rm{B}} = \pi / 2$, while the fit allows $T,B,\phi_{\rm{B}}$ to float; also included in the fit are the parameters to take into account the birefringence effects of the cell windows ($\theta_{\rm{BR}},\phi_{\rm{BR}}$). A fit without the effect of the cell window birefringence (broken line) is included for comparison.}
\label{fig:S3_fit}
\end{figure}

At the high magnetic field strength used for this work, optical rotation phenomena can provide additional information about the medium through which the light propagates~\cite{Ejlli2020}.
In the case of an atomic medium, measuring this optical rotation via the Stokes parameters, as mentioned in section \ref{sec:theory}, proves to be of natural interest for understanding the interactions between the atoms and the external magnetic field.
In this case, we have experimentally measured the dichroism and birefringence of the atomic medium in the three bases corresponding to the $S_{1},S_{2}\textrm{ and }S_{3}$ parameters as shown in figures~\ref{fig:S1_fit},~\ref{fig:S2_fit} and \ref{fig:S3_fit}, respectively.

In order to carry out these measurements a set of polarising optics (PBS$+\lambda/2$, $\lambda/4$) and two photodiodes was set up, as seen in figure~\ref{fig:setup}, to measure the light transmitted through the experiment cell in terms of the linear detuning of the laser scan.
This allows for two orthogonal polarisation components to be recorded simultaneously and then be processed into the corresponding Stokes parameter for the basis in question.
We take the definitions of the Stokes parameters as used in references~\cite{Zentile2015b,Keaveney2017a}.
Figure~\ref{fig:S1_fit} shows the $S_{1}$ parameter, taken as the difference between the orthogonal linear polarisations (\textit{i.e.} horizontal and vertical) (eq.~\ref{eqn:S1}).
Figure~\ref{fig:S2_fit} shows the $S_{2}$ parameter, defined as the difference between the linear polarisations in a basis rotated by $\pi/4$, giving diagonal components in a Cartesian basis (eq.~\ref{eqn:S2}).
Lastly, figure~\ref{fig:S3_fit} shows the last Stokes parameter, $S_{3}$, as the difference between orthogonal circular polarisations in the helicity basis, (\textit{i.e.} left-hand and right-hand circular) (eq.~\ref{eqn:S3}).
In these cases, given a well-defined input polarisation, the transmitted light gives information regarding the linear and circular birefringence of the atomic medium.

Using our theoretical model, \emph{ElecSus}, we proceed to fit the data to each of the three Stokes parameters mentioned above.
\textcolor{black}{Very good agreement~\cite{Hughes2010} between the model (solid purple curve) and data (blue circles) is evident in figures~\ref{fig:S1_fit}, \ref{fig:S2_fit} and~\ref{fig:S3_fit}}.
Despite this, there are still slight discrepancies between the experimental data and the fit.
We can try and remove some of these errors by taking into account the birefringence of the windows of our experiment cell.
To do this, we include in our model the effects of two thin, birefringent windows interacting with the electric field of our laser beam twice: once before the light enters the atomic medium and once when the light has passed through the atomic medium and exits the cell.
We carry out these calculations by using the Jones matrix formalism \cite{Keaveney2017a}, so that in this case the transmitted electric field $E_{\rm{out}}$ in our experiment can be written as
\begin{eqnarray*}
	E_{\rm{out}} = \mathcal{M}_{\theta_{\rm{BR}},\phi_{\rm{BR}}} \times \mathcal{J}_{\rm{atoms}} \times \mathcal{M}_{\theta_{\rm{BR}},\phi_{\rm{BR}}}\times E_{\rm{in}},
	\label{eq:biref}
\end{eqnarray*}
where $E_{\rm{in}}$ is the incident electric field, $\mathcal{M}_{\theta_{\rm{BR}},\phi_{\rm{BR}}}$ is the Jones matrix representing the birefringent window of the cell and $\mathcal{J}_{\rm{atoms}}$ is the Jones matrix representing the dichroic and birefringent atomic medium.
The matrix $\mathcal{M}_{\theta_{\rm{BR}},\phi_{\rm{BR}}}$ has been included twice to account for the entry and exit windows of the experiment cell.
This output electric field can be multiplied by the appropriate Jones' matrices to give the desired polarisation components to process into the form of the different Stokes' parameters.

From our fits we can see that the birefringence due to the cell windows is considerably small.
In this case the cell windows have a thickness of $300~\rm{\mu m}$ each \cite{Knappe2005}.
Using the literature value for the Verdet coefficient of glass \cite{Carr2020} at the wavelength of the Rb D2 line ($\sim 780$~nm), each cell window induces a rotation of $\approx 0.05^{\circ}$.
Here we make an initial assumption that both of these windows are identical in their birefringent properties.
Figures~\ref{fig:S1_fit},~\ref{fig:S2_fit} and~\ref{fig:S3_fit} show \textcolor{black}{very good agreement between the data (blue circles) and the model considering birefringence (broken green curve)}~\cite{Hughes2010}.
The residuals R shown in these figures correspond to the results of the fits that include the birefringence of both cell windows.
We obtain average values of $(0.96\pm0.16)$ radians ($(55\pm9)^{\circ}$) for the angle $\theta_{\rm{BR}}$ and of $(0.06\pm0.03)$ radians ($(3\pm2)^{\circ}$) for the angle $\phi_{\rm{BR}}$.
It is worth noting that these values correspond to a fit of the effect \emph{both} cell windows have on the electric field transmitted through the cell.
Comparing the fits to the experimental data with and without the birefringence effects from the cell windows we can see that this effect is particularly evident in the $S_{3}$ parameter, as seen in figure~\ref{fig:S3_fit}.
Due to the definition of said parameter (eq.~\ref{eqn:S3}) we can proceed to say that the cell windows have a predominantly circular birefringence.
The discrepancy between the rotation induced in the cell due to the field ($\approx 0.1^{\circ}$) and the measured value for $\phi_{\rm{BR}}$ is assumed to be caused by the heating of the windows as well as mechanical stresses from the fabrication process and the optical setup.
In particular, we exploit the sensitivity of the atomic system to optical rotation, in this case in the basis of orthogonal circular polarisation states, to obtain a signal that enhances these effects so that they are clearly visible.
As such, this experimental system provides a tool to characterise these birefringence effects due to vapour cell windows in order to reduce systematic errors in future measurements. 

\section{Conclusions}
\label{sec:conclusions}
\textcolor{black}{In this work we have presented a spectroscopy-based technique using a thermal alkali-metal vapour that allows absolute magnetic field strength and orientation of the field with respect to the light polarisation to be measured. 
An isotopically enriched sample of $^{87}$Rb was used; however the principles behind this technique can be easily applied to other alkali-metal atoms}.
\textcolor{black}{Very good agreement was found between the theoretical model used and our detailed spectroscopic measurements of the Stokes parameters of the medium.}
We have used polarisation-sensitive detection in order to better constrain the polarisation angle measured, as well as measure the birefringence effects due to the vapour cell windows.
Using this technique it is also possible to envisage a precise spectroscopy setup for atomic magnetometry in large ($>1$~T) fields.
Furthermore, the work here presented opens up new areas of research using atomic vapours, such as measurement of fundamental constants via precision thermometry using the ground-state populations.

\section{Acknowledgements}

The authors would like to thank Jacques Vigu\'e for kindly providing the permanent magnet used, James Keaveney for his contributions and insightful discussions, Danielle Pizzey for her comments and suggestions on the manuscript, Stephen Lishman and members of the \textcolor{black}{mechanical }workshop for their aid in fabricating the mechanical components for the experiment.
\textcolor{black}{The authors} acknowledge funding from EPSRC (Grant  EP/R002061/1) and Durham University.
FSPO gratefully acknowledges a Durham Doctoral Scholarship.
The data presented in this paper are available from \footnote{\url{DOI://10.15128/r1kk91fk595}}.

\bibliography{1.5T_Spectroscopy_Paper.bib}

\end{document}